# A study on deep feature extraction to detect and classify Acute Lymphoblastic Leukemia (ALL)

Sabit Ahamed Preanto

4IR Research Cell

Daffodil International University, Dhaka, Bangladesh

preanto15-5059@diu.edu.bd

Dr. Md. Taimur Ahad

4IR Research Cell

Daffodil International University, Dhaka, Bangladesh

taimurahad.cse@diu.edu.bd

Yousuf Rayhan Emon

4IR Research Cell

Daffodil International University, Dhaka, Bangladesh

yousuf15-3220@diu.edu.bd

Sumaya Mustofa

4IR Research Cell

Daffodil International University, Dhaka, Bangladesh

sumaya15-3445@diu.edu.bd

Md Alamin

4IR Research Cell

Daffodil International University, Dhaka, Bangladesh

amin15-4230@diu.edu.bd

## 1. Abstract

Acute lymphoblastic leukaemia (ALL) is a blood malignancy that mainly affects adults and children. This study looks into the use of deep learning, specifically Convolutional Neural Networks (CNNs), for the detection and classification of ALL. Conventional techniques for ALL diagnosis, such bone marrow biopsy, are costly and prone to mistakes made by hand. By utilising automated technologies, the research seeks to improve diagnostic accuracy. The research uses a variety of pre-trained CNN models, such as InceptionV3, ResNet101, VGG19, DenseNet121, MobileNetV2, and DenseNet121, to extract characteristics from pictures of blood smears. ANOVA, Recursive Feature Elimination (RFE), Random Forest, Lasso, and Principal Component Analysis (PCA) are a few of the selection approaches used to find the most relevant features after feature extraction. Following that, machine learning methods like Naïve Bayes, Random Forest, Support Vector Machine (SVM), and K-Nearest Neighbours (KNN) are used to classify these features. With an 87% accuracy rate, the ResNet101 model produced the best results, closely followed by DenseNet121 and VGG19. According to the study, CNN-based models have the potential to decrease the need for medical specialists by increasing the speed and accuracy of ALL diagnosis. To improve model performance, the study also recommends expanding and diversifying datasets and investigating more sophisticated designs such as transformers. This study highlights how well automated deep learning systems do medical diagnosis.

## 2. Introduction

Data is very important for machine learning applications (Ahad et al., 2024; Emon et al. 2024, Mustofa et al. 2024). Acute Lymphoblastic Leukemia (ALL) is a fatal form of leukemia which causes a high mortality rate among children and adults (Saeed et al., 2024; Attallah 2024). ALL is a type of blood cancer which occurs due to excessive white blood cell production in our body (Anwar & Alam 2020). At the beginning, it originates in the lymphatic system which start in the bone marrow and can rapidly spread across various parts of the body (Aziz et al., 2024). In within weeks, it can circulate in children's body and take the life (Attallah 2024; Rahman et al., 2023). To detect ALL, a bone marrow biopsy is performed, where an experienced pathologist conducts a microscopic analysis of blood sample. The procedure is very costly and may lead to inaccurate detection due to human errors (Saeed et al., 2024).

Machine learning has proven very effective in biomedical engineering (Paul et al., 2022; Shefat et al., 2023). To improve accuracy in classifying ALL images, several modern machine learning approaches has been used along with deep learning approaches. Several automated approaches have been developed to reduce dependency of health care professionals in detecting microscopic images of ALL. By using computer-assisted diagnosis, even inexperienced healthcare workers can detect ALL accurately which eventually will save time and money (Mohammed et al., 2023; Attallah 2024).

## 3. Background Study

Recently, Saeed et al. (2024) proposed DeepLeukNet, an automated system for diagnosing Acute Lymphoblastic Leukemia disease using a convolutional neural network technique attaining 99.61% accuracy. To effectively diagnose acute lymphoblastic leukemia from blood smear pictures, a new analysis technique with machine learning techniques and a composite learning approach were proposed by Bose & Bandyopadhyay (2024) which achieved 99.9% accuracy with a hybrid model of SVM and ResNet50. Al-Bashir et al. (2024) used CNN-based algorithms like AlexNet, DenseNet, ResNet, and VGG16 to detect leukemia and classify its types which gained accuracies, reaching 99.8%, 99.7%, and 94% for training, validation, and testing sets, respectively. Rehman et al. (2018) proposed a CNN-based robust segmentation and deep learning methods to train the model on the bone marrow images that achieved 97.78% accuracy in ALL diagnosis. MoradiAmin et al. (2024) motivated to develop accurate and efficient automatic system to distinguish ALL cells from similar lymphocyte subtypes without the need for direct feature extraction where they enhanced the microscopic images using histogram equalization and then a fuzzy C-means clustering algorithm is employed to segment cell nuclei. Finally, they have trained CNN using labeled dataset and the performance was evaluated using VGG-16, DenseNet, and Xception achieving 97% accuracy. An improved ResNet50 CNN model using a hybridization of Particle Swarm Optimization (PSO) to detect ALL and its subtypes is proposed by Özbay et al., (2023). By removing the last 4 layers from the ResNet50 backbone model and adding 11 new layers, the model has been trained on an augmented dataset and the result outperforms 99.65% accuracy. Ghaderzadeh et al. (2022) made a publicly available ALL dataset containing 3562 PBS images from 89 patients suspected of ALL, including 25 healthy individuals with a benign diagnosis (hematogone) and 64 patients with a definitive diagnosis of ALL subtypes. They employed EfficientNet, MobileNetV3, VGG-19, Xception, InceptionV3, ResNet50V2, VGG-16, NASNetLarge, InceptionResNetV2, and DenseNet201 CNN models for feature extraction and among all models, DenseNet201 achieved best performance gaining 99.85% accuracy, 99.52% sensitivity, and 99.89% specificity. Surya Sashank et al. (2021) presented two different classification models for detection of ALL utilizing ALL-IDB2 dataset of microscopic images of blood. They used AlexNet based detection for image pre-processing after feature extraction and classification which achieved 100% accuracy in classification. Alam, A., & Anwar, S. (2021) discussed a computer-aided automated diagnosis system for detection of ALL using deep-learning models which was done by a pretrained AlexNet model. The experiment was done using microscopic blood cell images for detecting malignant cells. The method achieved an accuracy of 98% without using image segmentation technique or feature extraction technique. Anwar, S., & Alam, A. (2020) performed various data augmentation techniques to train a CNN model on 515 images with automated feature extraction which gained an average accuracy of 95.54% during training and 99.5% average accuracy during testing.

This group of researchers mainly focused on developing and implementing various types of CNN architectures for detection and classification of ALL. They used various types of pre-trained models such as AlexNet, RestNet, MobileNetV3, VGG-19 and other type of custom designed CNNs which achieved accuracy ranging from 95%-100% which demonstrate the effectiveness of CNNs in ALL detection and classification. Although their study shows promising results, many studies have big gaps between training and testing performance. Moreover the models could be implemented on larger and more diverse datasets to ensure viability and some studies also has lacking in exploring multi-class problems or other leukemia types to broaden the implementation of the models.

Attallah (2024) used a novel CAD using wavelet-based CNNs for ALL detection and subtype classification without pre-segmentation which exceeded 100% accuracy in both ALL detection and subtype classification with just 88 and 146 features. Hassan et al. (2024) introduced with two novel architectures, in the first approach they used Knowledge Nevestro Classification (KNC) leveraging two teacher models, DenseNet and ResNet (50) and in the second approach they used Nevestro DenseNet Attention (DAN) which represented a variant of DenseNEt with Attention (NDA) architecture which had an impressive metrics with accuracy reaching 0.991. An ensemble strategy using CNN-GRU-BiLSTM and MSVM classifier was proposed by Mohammed et al. (2023) to detect ALL cells in three stages, where at first they implemented image pre-processing through oversampling process then generated deep spatial features using CNN and utilized gated recurrent unit (GRU)-bidirectional long short-term memory (BiLSTM) architecture to extract long-distance dependent features. Finally, classification was done using multiclass support vector machine (MSVM) and a softmax function. The DenseNet-201 model gained accuracy of 96.29% and 96.23% F1-score using MSVM classifier. Rahman et al., (2023) extracted features from individual images with CNN models then transferred features to Principal Component Analysis (PCA), Linear discriminant Analysis (LDA) and SVC Feature Selectors along with two nature inspired algorithms like Particle Swarm Optimization (PSO) and Cat Swarm Optimization (CSO) and then finally the model was assessed which gained 99.84% accuracy by integrating ResNet50 CNN architecture, SVC feature selector, and LR classifiers. A ViT-CNN ensemble model was introduced by Jiang et al. (2021) to classify cancer cells images and normal cells images in the diagnosis of ALL and the following classification model reached 99.03% accuracy on test set. Rodrigues et al. (2022) proposed a hybrid model using a genetic algorithm (GA) and a residual convolutional neural network (CNN) and finally ResNet-50V2 to predict ALL using microscopic images in ALL-IDB dataset. In the following study GA hyperparameters lead to highest accuracy obtaining 98.46%.

This group of researchers mainly worked on ensemble and hybrid models, which is a combination of different machine learning and deep learning techniques. The combination of ensemble and hybrid model had an outstanding accuracy of 99%. But to improve the performance they could have explored more recent transformer-based architectures. The model was trained on competitively smaller datasets which and some studies has explored more complex models which has led to difficulties in their decision-making process.

Masoudi, B. (2023) proposed a three-stage model based on transfer learning called variable-kernel channel-spatial attention (VKCS) to provide a more accurate diagnosis of ALL. They used EfficientNet-V2M for feature extraction and gained 100% accuracy in VKCS model for ALL-IDB1 dataset and 99.6% accuracy for the ALL-IDB2 dataset. Deep Convolutional Neural Networks (DNNs) was used by Anilkumar et al., (2022) to classify ALL from an online image bank of American Society of Haematology (ASH). A classification accuracy of 94.12% is achieved by the study in isolating the B-cell and T-cell ALL images using a pretrained CNN AlexNet as well as LeukNet, a custom-made deep learning network designed by the proposed work. Rezayi et al., (2021) employed ResNet50 and VGG-16 on a collected dataset from CodaLab competition to classify leukemic cells from normal cells in microscopic images. Six common machine learning techniques were used to classify ALL validation where achieved 84.62% accuracy with VGG-16 and 81.72% accuracy with Random Forest. And lastly five CNN algorithms were implemented by Sunil, S., & Sonu, P. (2021) to classify and the cancerous and non-cancerous cells. The system accepts the blood cell image from the user and predicts whether the cell contains one/more blasts depending upon the prediction value obtained from the CNN algorithm on the stained cell image.

This group implemented specialized techniques while focusing on developing novel architectures to achieve the most effective approach for ALL detection and classification. Although some of the studies has gained 100% accuracy, some reported lower accuracies, and some studies didn't mention any accuracy. The researchers could improve their works by providing more quantitative results. Some researchers could be improved by exploring recent transformer based architectures to enhance performance.

## 4. Methodology

*4.1 Dataset Collection*

Acute Lymphoblastic Leukemia (ALL) dataset has been collected from Kaggle (kaggle.com/datasets/mehradaria/leukemia/data) which had a total 3256 images from 89 patients suspected of ALL, including 25 healthy individuals with a benign diagnosis (hematogone) and 64 patients with a definitive diagnosis of ALL subtypes, Early Pre-B, Pre-B, and Pro-B ALL. The captured images are microscopic views of peripheral blood smear samples, which serve as primary medium for ALL detection.

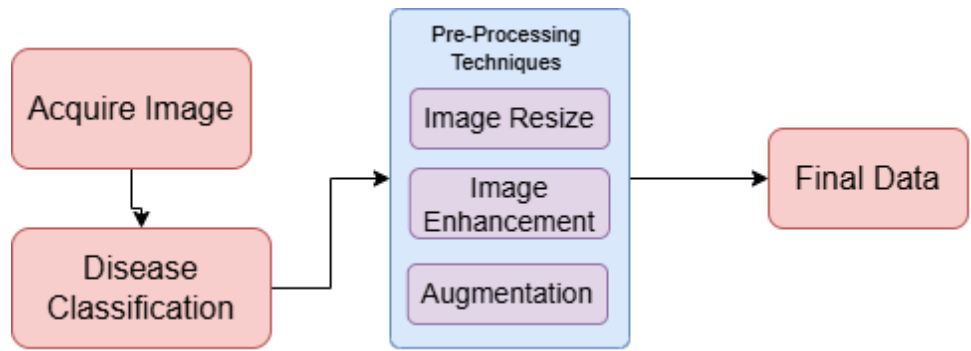

Figure 4.1: Data Preparation Steps

*Table 4.1: Dataset Description*

| Name of the class | Image of the disease | Number of images per class |
|---|---|---|
|  |  |  |
| Benign | 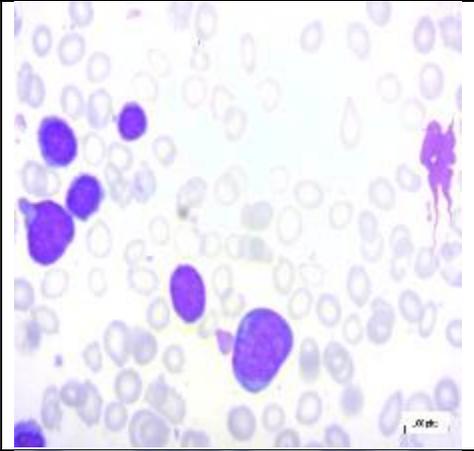 | 504 |
| Early | 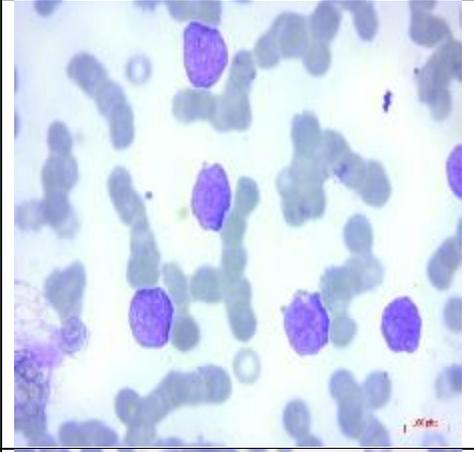 | 985 |
| Pre | 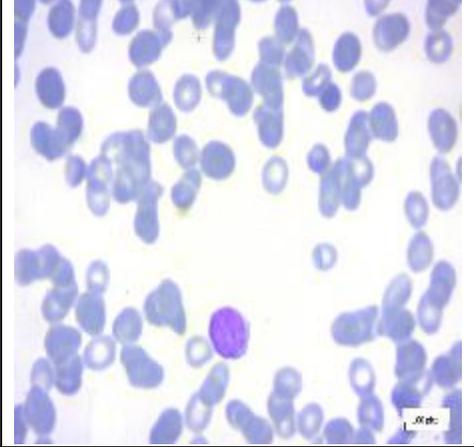 | 963 |
| Pro | 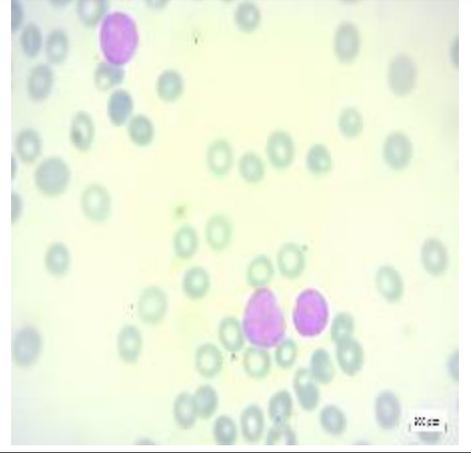 | 804 |
| | Total Images | 3256 |

*4.2 Image preprocessing techniques:*

At first, we processed the images by resizing them to a uniform resolution (320x240 px) to standardize input dimension. Then we employed histogram equalization to enhance contrast which allowed for more accurate feature detection. Finally, we have employed data augmentation in "Benign" class as it had less data compared to other classes.

*4.3 Feature extraction*

We have used five pre-trained convolutional neural networks (CNNs) to extract features from processed images. Each model was chosen based on their effectiveness in handling image classification tasks for detecting ALL. The models are: ResNet101, VGG19, InceptionV3, DenseNet121, MobileNetV2. We have removed the final fully connected layer using the output of the preceding convolutional layers for all models. For this, deep features were provided representing high-level abstractions from the images.

*4.4 Feature selection & classification*

We have applied five feature selection methods to reduce the dimensionality of the extracted features and enhance classification performance. The feature selection methods were:

ANOVA: Selected features based on the highest variance, selecting 500 features.

Recursive Feature Elimination (RFE): Utilized to identify the top 200 features that contributed most significantly to model performance.

Random Forest: Selected important features by evaluating feature importance scores, resulting in a variable number of selected features.

Lasso (Least Absolute Shrinkage and Selection Operator): Applied to enforce sparsity, selecting a compact set of relevant features (13 to 885 features depending on the model).

Principal Component Analysis (PCA): Reduced dimensionality by selecting 512 components to retain most of the data variance.

To assess the performance of selected features, four machine learning classifiers- K-Nearest Neighbors (KNN), Support Vector Machine (SVM), Random Forest, Naïve Bayes were applied. Each classifier was trained on the selected features and evaluated based on their ability to classify the images into one of the four ALL subtypes (Early Pre-B, Pre-B, Pro-B) or the benign class. The model was measured based on accuracy, precision, recall, and F1 score.

*4.5 Experimental Setup:*

The experiment was conducted on google collaboratory where we used NVIDIA TPU for computation and python was used as programming language. The whole dataset was divided into 80% training and 20% testing to ensure fair evaluation of model performance. The attention-based transformer and conventional CNN: 72% for training, 18% for validation, and 10% for testing used in the distribution. The following ratios assure a balanced data distribution for reliable training, validation, and assessment.

*4.6 Experimental Procedure:*

The following experiment is divided into two phases. In phase one, feature selection was performed using five CNN models and feature selection methods. In phase two, five machine learning classifiers were used in the selected features. Finally, a voting ensemble method was employed to combine the prediction from all models to improve classification accuracy.

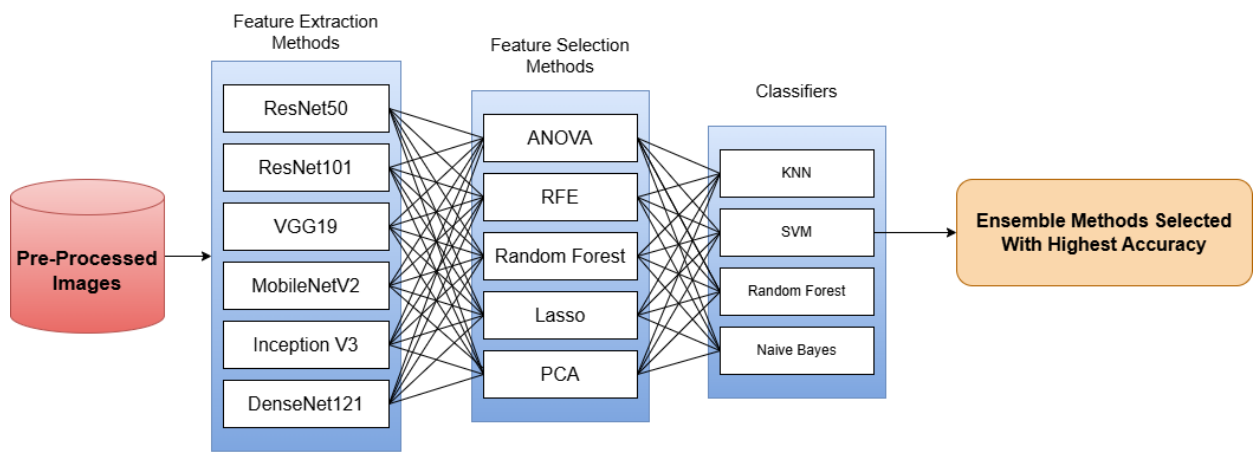

Figure 4.2: Feature Extraction, Selection and Classifier selection Methodology

*4.7 Statistical Analysis:*

The data normalization step scaling was applied to keep the features within a definite range. To normalize the features "MinMax" scaler was used to convert all features into the range [0,1], where 0 and 1 represent the minimum and maximum values of a feature or variable, respectively.

$$Xscaled = \frac{X-Xmin}{Xmax-Xmin}$$

To extract features from the extracted dataset of image extraction methods, five feature selection methods-ANOVA, RFE, Random Forest, Lasso, and PCA were used. Finally, four ML classifiers (KNN, SVM, RandomForest, and Naïve Bayes) were used to classify the diseases. The full experiment is demonstrated in the below diagram.

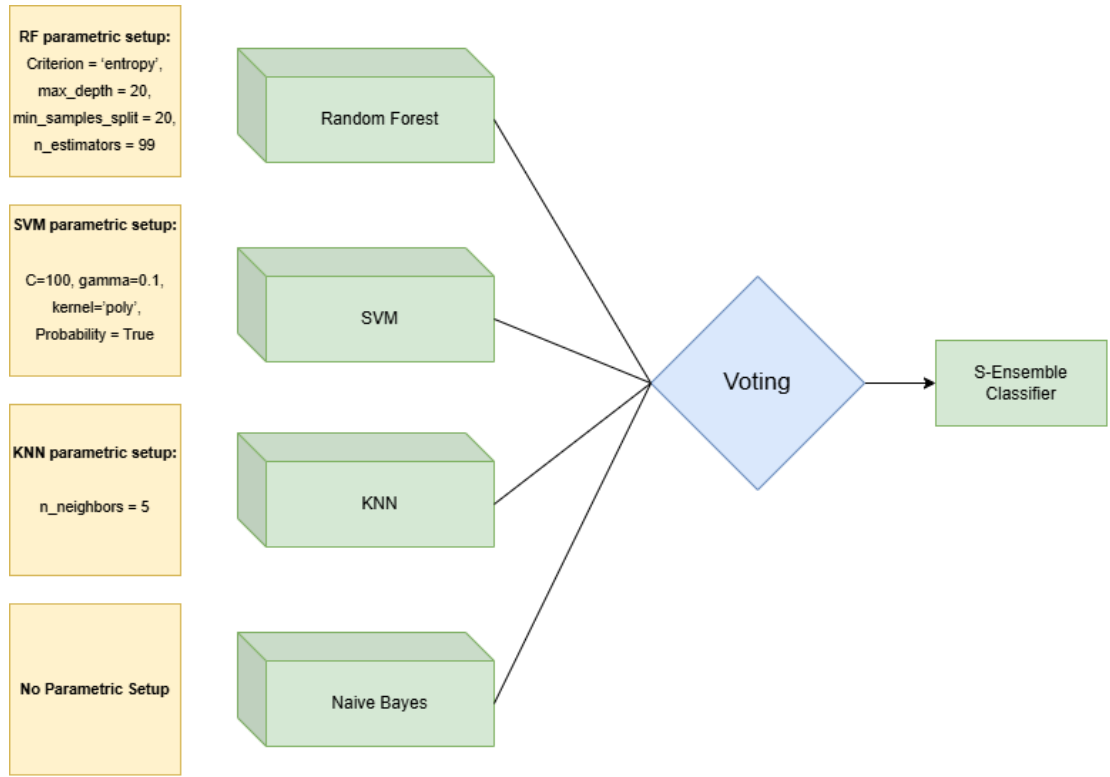

Figure 4.3: Proposed S-Ensemble Classifier

*Performance Measurement Techniques:*

The performance of this study has been measured based on a few criteria such as:

$$Accuracy = \frac{TP + TN}{TP + FP + TN + FN}$$

$$Precision = \frac{TP}{TP+FP}$$

$$Recall = \frac{TP}{TP+FN}$$

$$F1-score = \frac{2 \times Precision \times Recall}{Precision + Reacll}$$

## 5. Results and analysis

In this section, we provide an in-depth analysis of the results achieved from the feature extraction models, ML classifiers and the voting ensemble method for detecting ALL.

*5.1 Performance Measurement (Feature extraction models)*

The overall performance of the feature extraction methods ResNet101, VGG19, MobileNetV2, Inception V3, and DenseNet121 is shown in Tables 2, 3, 4, 5, and 6. To select the best features, five feature selection methods (ANOVA, RFE, Random Forest, Lasso, PCA) were implemented and after selecting relevant features classification algorithms (Random Forest, KNN, SVM, Naïve Bayes) were applied to determine the performance of each feature selector and classifier.

*5.1.1 Performance of ResNet101*

Table 5.1 represents the extracted features by ResNet101. Among all feature selection techniques, SVM showed the overall best result when paired with different feature selection methods. With PCA (512) and Lasso (13), SVM achieves the highest accuracy around (86.3%-87%) and F1-scores around 0.85-0.87. The K-NN model also showed reliable performance across all feature selection methods as it maintained a steady accuracy of around 85.3%. Random Forest had a variability in results and achieved best performance with ANOVA(500) and RFE(200) achieving around 82.8% accuracy. Naïve Bayes ranked as the weakest among all the classifiers with accuracies below 80% and lower F1-score comparing to other classifier models.

*Table 5.1: Overall Performance of ResNet101*

| Feature Selection Method | Classification Algorithm | Performance Matrix | | | |
|---|---|---|---|---|---|
| | | Accuracy | Precision | Recall | F1-score |
| ANOVA(500) | K-NN | 0.853 | 0.85 | 0.853 | 0.849 |
| | SVM | 0.845 | 0.842 | 0.845 | 0.843 |
| | Random Forest | 0.819 | 0.851 | 0.819 | 0.764 |
| | Naïve Bayes | 0.798 | 0.816 | 0.798 | 0.796 |
| RFE(200) | K-NN | 0.856 | 0.853 | 0.856 | 0.852 |
| | SVM | 0.833 | 0.829 | 0.833 | 0.83 |
| | Random Forest | 0.828 | 0.86 | 0.828 | 0.783 |
| | Naïve Bayes | 0.822 | 0.842 | 0.822 | 0.821 |
| Random Forest(221) | K-NN | 0.853 | 0.849 | 0.853 | 0.849 |
| | SVM | 0.853 | 0.85 | 0.853 | 0.851 |
| | Random Forest | 0.814 | 0.858 | 0.814 | 0.755 |
| | Naïve Bayes | 0.784 | 0.802 | 0.784 | 0.777 |
| Lasso(13) | K-NN | 0.851 | 0.848 | 0.851 | 0.849 |
| | SVM | 0.87 | 0.868 | 0.87 | 0.868 |
| | Random Forest | 0.814 | 0.882 | 0.814 | 0.749 |
| | Naïve Bayes | 0.761 | 0.771 | 0.761 | 0.752 |
| PCA(512) | K-NN | 0.845 | 0.841 | 0.845 | 0.842 |
| | SVM | 0.863 | 0.862 | 0.863 | 0.862 |
| | Random Forest | 0.794 | 0.671 | 0.794 | 0.721 |
| | Naïve Bayes | 0.672 | 0.705 | 0.672 | 0.683 |

*5.1.2 Performance of VGG19*

Table 5.2 represents the extracted features by VGG19, where SVM consistently outperforms all other classifiers across all feature selection methods, with accuracy reaching 84.2% using RFE (200) and 83.1% with Random Forest(221) gaining strong precision and F1-scores around 0.83. K-NN showed moderate performance achieving its highest accuracy of 79.1% with RFE(200) and

77% with Random Forest(221) which is slightly lower compared to SVM. Random Forest exhibited decent performance across feature selection methods. Its best accuracy of 78.7% is recorded with RFE(200) and Random Forest(221) which is effective but still not as proficient as SVM in optimizing classification. Finally, Naïve Bayes underperformed significantly with its highest accuracy of 72.5% using RFE(200).

*Table 5.2 Overall Performance of VGG19*

| Feature Selection Method | Classification Algorithm | Performance Matrix | | | |
|---|---|---|---|---|---|
| | | Accuracy | Precision | Recall | F1 score |
| ANOVA(500) | K-NN | 0.758 | 0.743 | 0.758 | 0.745 |
| | SVM | 0.822 | 0.819 | 0.821 | 0.82 |
| | Random Forest | 0.782 | 0.769 | 0.782 | 0.735 |
| | Naïve Bayes | 0.595 | 0.62 | 0.595 | 0.568 |
| RFE(200) | K-NN | 0.791 | 0.779 | 0.791 | 0.779 |
| | SVM | 0.842 | 0.836 | 0.842 | 0.838 |
| | Random Forest | 0.787 | 0.771 | 0.787 | 0.745 |
| | Naïve Bayes | 0.725 | 0.732 | 0.725 | 0.72 |
| Random Forest(221) | K-NN | 0.77 | 0.752 | 0.77 | 0.753 |
| | SVM | 0.831 | 0.825 | 0.831 | 0.827 |
| | Random Forest | 0.787 | 0.774 | 0.787 | 0.744 |
| | Naïve Bayes | 0.722 | 0.73 | 0.722 | 0.717 |
| Lasso(13) | K-NN | 0.773 | 0.756 | 0.773 | 0.757 |
| | SVM | 0.813 | 0.808 | 0.813 | 0.81 |
| | Random Forest | 0.776 | 0.761 | 0.776 | 0.727 |
| | Naïve Bayes | 0.666 | 0.676 | 0.666 | 0.653 |
| PCA(512) | K-NN | 0.75 | 0.733 | 0.75 | 0.734 |
| | SVM | 0.825 | 0.821 | 0.825 | 0.823 |
| | Random Forest | 0.745 | 0.62 | 0.745 | 0.673 |

| | Naïve Bayes | 0.489 | 0.471 | 0.489 | 0.457 |

### 5.1.3 Performance of Inception V3

Table 5.3 represents the extracted features by InceptionV3, where SVM consistently outperformed all other algorithms in terms of accuracy, precision, recall, and F1 score. SVM gained its highest accuracy of 77.1% with Random Forest (221), followed closely by 75.2% with RFE(200). K-NN also performed decently with its best accuracy 70.6% under RFE (200). Naïve Bayes again underperformed with accuracy scores as low as 41.6% using PCA(512). Overall, SVM proves to be the most effective classifier when combined with RFE(200) and Random Forest (221).

*Table 5.3 Overall Performance of InceptionV3*

| Feature Selection Method | Classification Algorithm | Performance Matrix | | | |
|---|---|---|---|---|---|
| | | Accuracy | Precision | Recall | F1 score |
| ANOVA(500) | K-NN | 0.637 | 0.618 | 0.637 | 0.624 |
| | SVM | 0.738 | 0.726 | 0.738 | 0.72 |
| | Random Forest | 0.681 | 0.647 | 0.681 | 0.624 |
| | Naïve Bayes | 0.507 | 0.523 | 0.473 | 0.473 |
| RFE(200) | K-NN | 0.706 | 0.69 | 0.706 | 0.689 |
| | SVM | 0.752 | 0.741 | 0.752 | 0.739 |
| | Random Forest | 0.704 | 0.698 | 0.704 | 0.653 |
| | Naïve Bayes | 0.653 | 0.657 | 0.653 | 0.638 |
| Random Forest(221) | K-NN | 0.687 | 0.676 | 0.687 | 0.669 |
| | SVM | 0.771 | 0.769 | 0.771 | 0.77 |
| | Random Forest | 0.699 | 0.693 | 0.699 | 0.647 |
| | Naïve Bayes | 0.655 | 0.656 | 0.655 | 0.648 |
| Lasso(13) | K-NN | 0.689 | 0.675 | 0.689 | 0.67 |
| | SVM | 0.748 | 0.744 | 0.748 | 0.745 |

|  | Random Forest | 0.683 | 0.629 | 0.683 | 0.625 |
|  | Naïve Bayes | 0.61 | 0.674 | 0.61 | 0.57 |
| PCA(512) | K-NN | 0.65 | 0.636 | 0.65 | 0.634 |
|  | SVM | 0.747 | 0.739 | 0.747 | 0.742 |
|  | Random Forest | 0.655 | 0.54 | 0.655 | 0.591 |
|  | Naïve Bayes | 0.416 | 0.402 | 0.416 | 0.382 |

### 5.1.4 Performance of DenseNet121

The results for DenseNet121 across different feature selection methods are represented in Table 5.4. The highest accuracy was achieved by SVM with the Random Forest(221) gaining 85.4%. K-NN also showed promising results with highest accuracy of 82.2% with Random Forest (221) method. Random Forest also performed comparably well, reaching a peak accuracy of 82.1% under both the Random Forest (221) and Lasso (13) feature selection methods. However, its performance in precision and F1 score shows that the accuracy dropped to 79.1% with PCA (512). Naïve Bayes continued to struggle to deliver promising result with its best accuracy of 74.4% under RFE (200) and significantly low accuracy of 59.4% with PCA (512).

*Table 5.4 Overall Performance of DenseNet121*

| Feature Selection Method | Classification Algorithm | Performance Matrix | | | |
| --- | --- | --- | --- | --- | --- |
|  |  | Accuracy | Precision | Recall | F1 score |
| ANOVA(500) | K-NN | 0.807 | 0.796 | 0.807 | 0.8 |
|  | SVM | 0.845 | 0.84 | 0.845 | 0.842 |
|  | Random Forest | 0.827 | 0.838 | 0.827 | 0.8 |
|  | Naïve Bayes | 0.707 | 0.738 | 0.707 | 0.692 |
| RFE(200) | K-NN | 0.814 | 0.807 | 0.814 | 0.81 |
|  | SVM | 0.847 | 0.842 | 0.847 | 0.838 |
|  | Random Forest | 0.819 | 0.827 | 0.819 | 0.79 |
|  | Naïve Bayes | 0.744 | 0.789 | 0.744 | 0.735 |
| Random Forest(221) | K-NN | 0.822 | 0.812 | 0.822 | 0.816 |
|  | SVM | 0.854 | 0.85 | 0.854 | 0.85 |

| | Random Forest | 0.821 | 0.831 | 0.821 | 0.792 |
| | Naïve Bayes | 0.732 | 0.764 | 0.732 | 0.719 |
| Lasso(13) | K-NN | 0.817 | 0.808 | 0.817 | 0.811 |
| | SVM | 0.839 | 0.834 | 0.839 | 0.835 |
| | Random Forest | 0.821 | 0.836 | 0.821 | 0.786 |
| | Naïve Bayes | 0.683 | 0.721 | 0.683 | 0.662 |
| PCA(512) | K-NN | 0.811 | 0.801 | 0.811 | 0.805 |
| | SVM | 0.828 | 0.821 | 0.828 | 0.822 |
| | Random Forest | 0.791 | 0.673 | 0.791 | 0.719 |
| | Naïve Bayes | 0.594 | 0.63 | 0.594 | 0.595 |

### 5.1.5 Performance of MobileNetv2

Table 5.5 shows the performance of MobileNetV2 feature selection methods. Overall SVM consistently performed well across all feature selection methods. Among the feature selection methods, RFE(200) and Lasso(13) overall performed the highest accuracy gaining 81.1%. K-NN also showed promising performance with RFE (200) gaining the highest accuracy of 81%. PCA (512) appeared to be the least effective feature selection methods for Random Forest and Naïve Bayes algorithms.

*Table 5.5: Overall Performance of MobileNetv2*

| Feature Selection Method | Classification Algorithm | Performance Matrix | | | |
|---|---|---|---|---|---|
| | | Accuracy | Precision | Recall | F1 score |
| ANOVA(500) | K-NN | 0.794 | 0.782 | 0.794 | 0.787 |
| | SVM | 0.804 | 0.804 | 0.804 | 0.803 |
| | Random Forest | 0.788 | 0.776 | 0.788 | 0.764 |
| | Naïve Bayes | 0.646 | 0.662 | 0.646 | 0.621 |
| RFE(200) | K-NN | 0.81 | 0.799 | 0.81 | 0.803 |
| | SVM | 0.81 | 0.808 | 0.81 | 0.808 |
| | Random Forest | 0.791 | 0.784 | 0.791 | 0.773 |
| | Naïve Bayes | 0.689 | 0.703 | 0.689 | 0.667 |
| Random Forest(221) | K-NN | 0.799 | 0.786 | 0.799 | 0.791 |
| | SVM | 0.81 | 0.807 | 0.81 | 0.808 |
| | Random Forest | 0.784 | 0.769 | 0.784 | 0.761 |
| | Naïve Bayes | 0.681 | 0.702 | 0.681 | 0.655 |
| Lasso(13) | K-NN | 0.799 | 0.788 | 0.799 | 0.792 |
| | SVM | 0.811 | 0.807 | 0.811 | 0.809 |
| | Random Forest | 0.771 | 0.756 | 0.771 | 0.745 |
| | Naïve Bayes | 0.692 | 0.716 | 0.692 | 0669 |
| PCA(512) | K-NN | 0.784 | 0.772 | 0.784 | 0.775 |
| | SVM | 0.804 | 0.801 | 0.804 | 0.802 |
| | Random Forest | 0.745 | 0.622 | 0.745 | 0.674 |
| | Naïve Bayes | 0.528 | 0.584 | 0.528 | 0.527 |

*5.2 Performance of individual feature extraction models*

The study evaluated five pre-trained Convolutional Neural Networks (CNNs) for feature extraction: ResNet101, VGG19, InceptionV3. DenseNet121, and MobileNetV2. The performance metrics for each models are summarized below:

*Table 5.6 Performance comparison of the feature extraction methods*

| Accuracy | Number of extracted features | Accuracy (%) | Precision (%) | Recall (%) | F1-Score |
|---|---|---|---|---|---|
| ResNet101 | 2048 | 87 | 86.8 | 87 | 86.8 |
| VGG19 | 512 | 84.2 | 83.6 | 84.2 | 83.8 |
| DenseNet121 | 1024 | 85.4 | 85 | 85.4 | 85 |
| Inceptionv3 | 2048 | 77.1 | 76.9 | 77.1 | 77 |
| MobileNetv2 | 1280 | 81.1 | 80.7 | 81.1 | 80.9 |

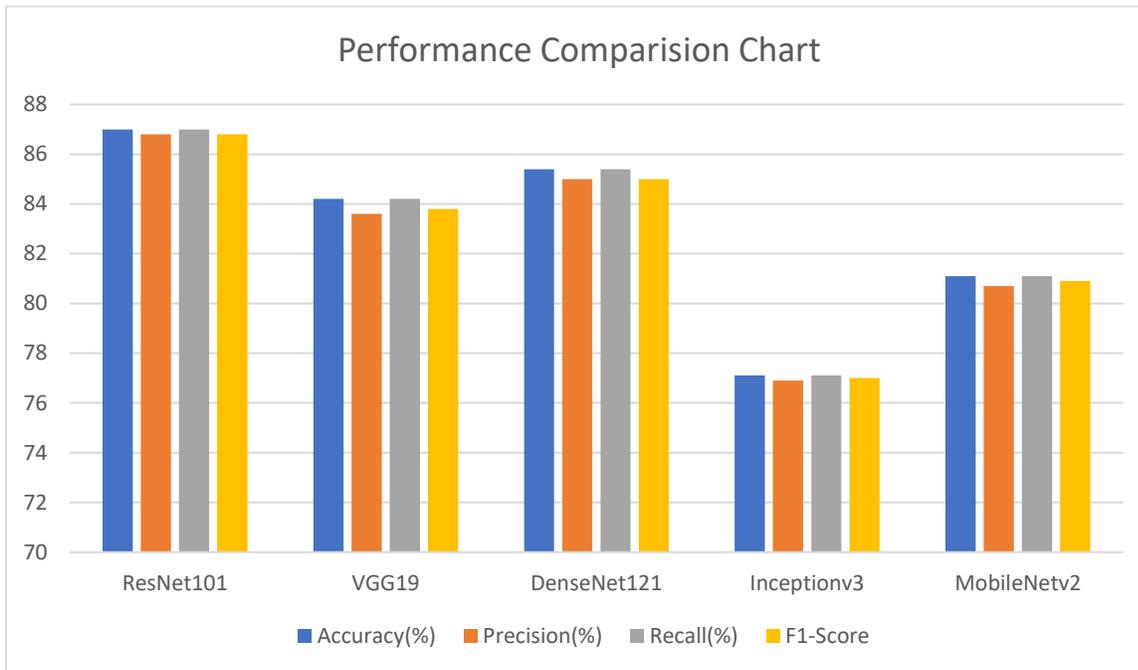

*Figure 5.1 Performance comparison of feature extraction models*

The following Figure 5.1 shows a performance comparison among all feature extraction methods, In the following chart, we can examine that ResNet101 consistently outperformed all other models in terms of accuracy, precision, recall and F1 score achieving an accuracy of 87%. DenseNet121 and VGG19 followed closely behind achieving accuracies of 85.4% and 84.2% respectively. InceptionV3 also showed promising performance with an accuracy of 77.1%,